# Resonant intersubband polariton-LO phonon scattering in an optically pumped polaritonic device

J-M. Manceau[1,*], N-L. Tran[1], G. Biasiol[2], T. Laurent[1], I Sagnes[1], G. Beaudoin[1], S. De Liberato[3], I. Carusotto[4], R. Colombelli[1,#]

[1] *Centre de Nanosciences et de Nanotechnologies, CNRS UMR 9001, Université Paris-Sud, Université Paris-Saclay, C2N - Orsay, 91405 Orsay cedex, France*
[2] *Laboratorio TASC, CNR-IOM, Area Science Park, S.S. 14 km 163.5 Basovizza, I-34149 Trieste, Italy,*
[3] *Department of Physics and Astronomy, University of Southampton, Highfield, Southampton, UK*
[4] *INO-CNR BEC Center and Dipartimento di Fisica, Universita di Trento, I-38123 Povo, Italy*

We report experimental evidence of longitudinal optical (LO) phonon-intersubband polariton scattering processes under resonant injection of light. The scattering process is resonant with *both* the initial (upper polariton) and final (lower polariton) states and is induced by the interaction of confined electrons with longitudinal optical phonons. The system is optically pumped with a mid-IR laser tuned between 1094 cm$^{-1}$ and 1134 cm$^{-1}$ ($\lambda$=9.14 µm and $\lambda$=8.82 µm). The demonstration is provided for both GaAs/AlGaAs and InGaAs/AlInAs doped quantum well systems whose intersubband plasmon lies at $\approx$10 µm wavelength. In addition to elucidating the microscopic mechanism of the polariton-phonon scattering, that is found to differ substantially from the standard single particle electron-LO phonon scattering mechanism, this work constitutes the first step towards the hopefully forthcoming demonstration of an intersubband polariton laser.

Since their first theoretical prediction two decades ago [1], intersubband (ISB) polaritons have been subject to abundant theoretical and experimental work. Shortly after their first experimental demonstration [2], it was understood that they offer the particularity to access a new regime of interaction, namely the ultra-strong coupling regime, which could initiate interesting quantum phenomena when modulated in time [3,4]. These theoretical predictions have been driving the experimental field for years towards the demonstration of unprecedented Rabi-splitting values [5-9]. The other key aspect of intersubband polaritons is their bosonic nature which, as for their excitonic counterpart [10], enables a regime of final state stimulation either *via* longitudinal optical phonons ($LO_{ph}$) scattering or *via* polariton-polariton scattering [11,12]. As a crucial new feature of ISB polariton devices, reference [13], that details a roadmap towards an ISB polariton laser, also points out that the upper density limit for bosonic behavior of ISB polaritons is not rigidly fixed - as is the case of the Mott density for excitons - but it can be engineered by design to a large extent with the electronic doping. This observation suggests that ISB polariton lasers should be relatively high output power devices.

To date, most - if not all - of the attempts to implement intersubband polaritonic light emitting devices were led under electrical pumping [14-19]. Such approach is hampered by the presence of the dark states of the ISB plasmon, that do not couple to the electro-magnetic field. The number of dark states being important, only a small fraction of the injected electrons tunnel into the *bright* polaritonic states and the efficiency of the process is dramatically reduced [20]. Optical pumping of intersubband polaritonic systems has instead concentrated on time domain studies demonstrating the ultrafast switching and bleaching of the strong coupling regime [21,22]. These experiments employ femtosecond-pulses tuned at the *interband* transition of the semiconductor quantum wells. These pulses induce carriers in the system - *via* electron-hole generation - but do not couple to the ISB polaritons. To date, and to the best of our knowledge, optical pumping directly at the intersubband polariton energies has never been reported. Yet this is the most effective manner to selectively inject energy into the bright polaritonic states.

Keeping in mind the ISB polariton laser as a long-term goal [13], in this letter we experimentally investigate the coupling between ISB polaritons and LO phonons. In particular, we report evidence of a scattering process that is resonant with both the initial (upper polariton, UP) and final (lower polariton, LP) states. This result has been enabled by pumping the system with a narrowband, resonant optical excitation at the UP energy. The

LO$_{ph}$ origin of the scattering mechanism is attested by the frequency of the scattered light that is red-shifted from the pump by 293.5 cm$^{-1}$ in a GaAs/AlGaAs quantum well (QW) system, and by 271 cm$^{-1}$ in an InGaAs/AlInAs QW system. Within the resolution limit of the measurement, the emitted/scattered light is spectrally as narrow as the pump laser, and - in agreement with its resonant character - its amplitude is proportional to the absorption of the injection state. The intensity of the scattered light is in fair agreement with theoretical calculations of the ISB polariton-LO phonon coupling along the lines of [11], which leads to an optimistic quantitative estimate for the laser threshold.

We have developed in recent years microcavity resonators that offer an energy minimum at *k=0* with a positive, parabolic dispersion for transverse magnetic (TM) polarized modes [23,24]. When coupled to an ISB plasmon, this permits to translate to the mid-IR spectral region the peculiar dispersion of excitonic polaritons, that has been the key enabling tool behind the demonstration of stimulated scattering and lasing in these systems. We have shown that this metal-insulator-metal geometry is operational over a very large frequency range and it is also compatible with electrical injection [19,25]. Most importantly, it suits optical pumping experiments as light can be injected at a different angle from the collection angle ($\theta=0$) hence minimizing stray light coming from the pump onto the detector. Figure 1a depicts the device under study. The top grating (period $\Lambda=4.26$ μm with filling factor of 72%) couples the impinging light by the surface and confines it within the active core volume. The metal-metal configuration allows maximum overlap of the electromagnetic field with the active core (sample HM3872) that consists in a 36-period repetition of 8.3-nm-thick GaAs QWs separated by 20-nm-thick Al$_{0.3}$Ga$_{0.7}$As barriers. The nominal surface doping per QW of the sample n$_{2D}$=4.4x10$^{12}$ cm$^{-2}$ is introduced as δ-layers in the center of the barriers. The absorption of the ISB transition at 300K and 78K has been already presented elsewhere [26].

The presence of the two new polaritonic eigenstates, characteristic of the strong coupling regime, is probed in the frequency domain using a Fourier Transform Infrared (FTIR) spectrometer. From the spectra recorded at different angles (from 13$^o$ to 61$^o$, in 2$^o$ steps, using a commercial, motorized angle reflection unit) the polaritonic dispersion can be readily obtained as explained in [23,24]. It is crucial to precisely assess the spectral position of the LP branch at normal incidence, as it is where the polaritons, optically injected in the upper branch, shall be scattered. We probed them using a different grating and electric field orientation: the reflectivity minima recorded at 300 K are marked as red stars along with the other angles (red dots) on the dispersion curve in Fig. 1b. The same measurement performed at 78 K is reported in Fig. 2a (red curves): it reveals as expected increase of the polariton branches separation, mainly governed by the blue shift of the UP by 30 cm$^{-1}$.

All the gathered information on the polaritonic dispersion of sample HM3872 permits to precisely identify the correct incidence angle to induce a polariton-LO phonon scattering process between the upper polariton injection and the lower polariton final states. From the literature [27], the LO$_{ph}$ energy at 78K and at $k=0$ is 294 cm$^{-1}$: this places the resonant pumping configuration at an incidence angle of 41 degrees, with a LP at 824 cm$^{-1}$ and an UP at 1118 cm$^{-1}$.

The experimental approach is schematically represented in Fig. 1a. The incident laser light is coupled in through an anti-reflection (AR) coated ZnSe lens (f #2) and the specular component of the reflection is measured using a power-meter. The scattered light is collected at normal incidence with an AR coated ZnSe lens (f #1.5) and analyzed with an FTIR spectrometer equipped with a liquid-nitrogen cooled Mercure Cadmium Telluride (MCT) detector. The pump laser source is a commercial quantum cascade laser (QCL) from Daylight solutions (MirCat system), tunable from 1026 to 1140 cm$^{-1}$ (step resolution of 0.1 cm$^{-1}$) with a maximum average output power in pulsed mode (100 kHz, 2µs) of 111 mW (peak power within the pulse is 5 times more). Using a knife-edge technique, we estimated the beam waist at the focal point to be 170 µm (1/e$^2$). The average estimated pump intensity can therefore reach up to 500 W/cm$^2$ (2 kW/cm$^2$ peak power).

It is possible to verify that the pump angle on the sample is correct. We record the 0$^{th}$ order reflected light from the sample while tuning the laser wavenumber in 2 cm$^{-1}$ steps. The reflectivity is obtained by dividing the data with the one recorded on a plane gold mirror. Figure 2a (blue curve) shows the reflectivity of the sample, across the pump laser tuning range, for an incidence angle of 41$^o$ and at 78 K: we can clearly identify the UP state. We therefore fix the pump wavelength at the maximum absorption of the injection (UP) state (1118cm$^{-1}$) and record, using a synchronous detection scheme, the scattered light collected at normal incidence for 77 mW of average power impinging on the sample. Given the 75% absorption of the sample at this angle and frequency, we estimate that 58 mW of average power are injected in the upper polariton state. To reduce the amount of stray light coming from the pump beam diffused at the surface of the sample, we have placed a long wave pass filter in front of our MCT detector (30 dB attenuation at the pump wavelength).

Figure 2b shows the experimental spectrum obtained in step-scan. One can clearly observe the presence of two peaks, one corresponding to the laser pump diffused by the imperfections of the grating, and a second lower peak separated by the energy of one $LO_{ph}$ (293 cm$^{-1}$) from the pump signal. The two peaks match with the maximum absorption of both injection and lower states. We have increased the resolution of the measurement down to 0.5 cm$^{-1}$ allowing a more precise measurement of the $LO_{ph}$ separation at 293.5 cm$^{-1}$. Interestingly, the scattered light is a replica of the injected laser light and no convolution with the $LO_{ph}$ lineshape is observable. Within this resolution limited scan, we can extract a full width at half maximum (FWHM) for the scattered light of 1 cm$^{-1}$.

We then study the dependence of the scattered light on the injection laser frequency. We have recorded the emission spectra for several different incident laser wavenumbers, in steps of 8 cm$^{-1}$. The pump intensity and integration time of the measurements were kept similar. The data are reported in Fig. 4. They reveal that the amount of scattered/emitted light is directly proportional to the absorption at the injection frequency, and that the $LO_{ph}$ energy separation is strictly respected as we change the pump wavenumber. Since we are collecting the scattered light within a large angular cone (±19°), the influence on the final state is weak, especially since the latter is relatively flat over this cone of angles and broad in frequency.

As a last piece of evidence of the $LO_{ph}$-related scattering mechanism, we have led the same experiments with a polaritonic sample based on the InGaAs/AlInAs system lattice matched on InP. The active core (sample InP1614) consists in 35-periods repetition of 10.5 nm thick InGaAs QWs separated by 15-nm-thick AlInAs barriers. The nominal doping of the sample $n_{3d}=1\times10^{18}$cm$^{-3}$ is introduced as bulk within the wells. From the literature, the $LO_{ph}$ in this system is expected to be around in the 270-274 cm$^{-1}$ range, depending on the exact alloy concentration [28]. Our measurements show a separation between the injected and scattered light of 271 cm$^{-1}$ at 78K, as shown in Fig. 5 that presents a typical spectrum, along with a measurement of the GaAs/AlGaAs sample. In order to highlight the difference in phonon separation for the two systems, we have plotted the frequency axis subtracting the pump frequency. One can clearly observe a difference of 22.5 cm$^{-1}$ between the spectral positions of the scattered light in the two systems. The full dispersion of the sample InP1614, as well the reflectivity of the final and injection states measured at 78 K, are available in the Supplementary Material [29].

These results have been obtained with an incident peak power of ≈ 1 kW/cm². This is an extremely low power with respect to the only *similar* literature available, that is the works on the quantum fountain laser (QFL) and Intersubband Raman Laser (IRL) reported at the end of the nineties [30,31]. These lasers, operating in the weak light-matter coupling regime, rely on population inversion within a 3 levels system present in asymmetrically coupled quantum wells. Of importance here is the observation that, in those works, no emission signal was ever detected for incident powers below approximately 1 MW/cm². Here, we are able to observe emission for incidence powers that are lower by at least three orders of magnitude. This is already a very interesting result. However, the important question is: how far are we from the onset of a stimulated process?

A quantitative estimate for the intensity of the scattered light can be obtained by deriving from the usual electron-phonon Frölich Hamiltonian the matrix element for the LO phonon-ISB plasmon coupling. A simple Fermi golden rule calculation generalizing the calculations in Ref. [32], predicts the following spontaneous, angle-integrated phonon scattering rate:

$$\Gamma_{ph} = \frac{\pi \omega_{LO} e^2 k_f L_{QW} f \left| u_{ISB}^i u_{ISB}^f \right|^2}{\hbar v_f^{gr} \epsilon_\rho} \quad (1)$$

where $\omega_{LO}$ is the LO phonon angular frequency, $k_{i,f}$ are the wave-vectors of the initial and final ISB polariton states, $v_f^{gr}$ is their group velocity (assuming an isotropic polariton dispersion around a minimum at $k=0$), $L_{QW}$ is the QW width, $u^{i,f}_{ISB}$ are the ISB plasmon Hopfield weights in the initial and final ISB polaritons, $\varepsilon_\rho = [\varepsilon_\infty^{-1} - \varepsilon_s^{-1}]^{-1}$ is the relative dielectric constant due to phonons, and *f* is a numerical factor of order 0.03 summarizing the intra-subband nature of the scattering process [32]. Inserting the actual figures for our device, we find a ~$10^{-7}$ branching ratio for phonon-polariton scattering compared to radiative and non-radiative losses. This value is in reasonable agreement with our experimental observations of the efficiency of the process. With an average power of 58 mW injected within the upper polaritonic state - based on the detector responsivity and the various losses induced by the measurement chain - we reach an estimated efficiency of 6x$10^{-9}$ for the measured scattered signal.

Formula (1) yields an ISB polariton/LO-phonon scattering time in the µs range, a value orders of magnitude longer than the ps-range scattering time of electron-phonon scattering typically found in QCL devices, i.e. the scattering of an electron from the second to the first subband in a semiconductor QW. This dramatic difference has a twofold origin. On one hand, it is partly a consequence of the much lower polariton mass compared to the electronic one. On the other hand, and most importantly, the single particle electron-phonon process in a QW is clearly of intersubband nature, while the the polariton-phonon scattering is essentially an intra-subband process, which also reduces the matrix element of the process. To our knowledge, it is the first time that this fundamental peculiarity of ISB polariton/LO phonon scattering has been highlighted.

To go beyond the spontaneous scattering rate and estimate the threshold for the onset of bosonic stimulation, we take advantage of our observation of a very long lifetime of the phonon mode and view the process as an upper ISB polariton being coherently down-converted into a signal-idler pair formed by a lower ISB polariton plus a phonon. Using the usual theory of parametric oscillation in planar devices [33], we can extract a value for the threshold

$$I_{th} = \frac{\hbar^2 \Gamma_{LO} \Gamma_{UP} \Gamma_{LP} \omega_{UP} \varepsilon_\rho}{4\pi |u_{ISB}^i|^2 |u_{ISB}^f|^2 \omega_{LO} f e^2 L_{QW}}, \quad (2)$$

Plugging into this equation actual values of our structure ($\hbar\Gamma_{up,}$~ 4 meV, $\hbar\Gamma_{lp}$~ 7meV and $\hbar\Gamma_{LO}$ <0.36 meV (LO-phonon Q-factor around 100) and 75% of absorbed light within the upper polariton state), we find a threshold intensity of the order of 70 kW/cm$^{2.}$, in agreement with the conclusions of Ref. [11] and, most remarkably, just a couple of orders of magnitude higher than one delivered on the sample with our actual pump system. With a reasonable focus as presented above, such level of intensities correspond to incident powers of ≈ 20W, that could be reached with pulsed $CO_2$ lasers or using energetic pulses emitted from an OPO system. Alternatively, pump-probe techniques could be used to populate the final state and facilitate the stimulated parametric scattering regime as elegantly demonstrated for exciton-polaritons [34]. An important difference compared to the excitonic case is however that the idler mode consists here of a phonon rather than another polariton, which makes the development of a coherently emitting device even more intriguing.

In conclusion, we have demonstrated spontaneous resonant ISB polariton - LO phonon scattering in an optically pumped device operating in the strong coupling regime between light and matter. Both the spectral position and the intensity of the scattering process are consistent with the theoretical predictions for GaAs and InGaAs based systems. Although the intra-subband nature of the electron-phonon process underlying the observed polariton-phonon scattering dramatically quenches its rate as compared to standard electron-phonon processes in QCLs, we predict an optimistic value in the 70 kW/cm$^2$ for the polariton lasing threshold intensity. We finally anticipate that the laser device might have practical application as a coherent source not only of Mid-IR light, but also of LO phonons.

We thank S. Pirotta, A. Bousseksou and F. Julien for useful discussions. We acknowledge financial support from the European Union FET-Open grant MIR-BOSE 737017, and from the European Research Council (IDEASERC) ("GEM") (306661). This work was partly supported by the French RENATECH network.

# References

*E-mail: jean-michel.manceau@u-psud.fr  
#E-mail: raffaele.colombelli@u-psud.fr

**Figure 1:** (a) Schematic representation of the experimental geometry. (b) Experimentally recorded reflectivity minima of the polaritonic sample (HM3872) at 300K. The two polaritonic branches are clearly separated and the reflectivity minima at 0° degree are marked with stars as they required a different experimental configuration to be measured.

**Figure 2**: (a) Spectral reflectivity of the sample at normal incidence (red lines) and at 41° incidence (blue line). The LP state evident in the red line corresponds to the final state, while the UP state in the blue line corresponds to the injection state. Both measurements were performed at 78 K. The injection state (blue line) is recorded step by step using the QCL and its specular reflection. The final state (red line) is recorded within the FTIR. (b) Emission spectrum of sample HM3872 under resonant light injection. The laser incident power is 58 mW. The measurement was performed in step-scan configuration at a resolution of 4cm$^{-1}$ and a temperature of 78K. One can clearly observe the correspondence with the injection and final state, separated by one LO$_{ph}$ .

**Figure 3**: Emission spectra of sample HM3872 under resonant optical pumping recorded with increasing spectral resolutions (from the top: 4 cm$^{-1}$, 1 cm$^{-1}$, 0.5 cm$^{-1}$). Within the resolution limit of the spectrometer, it allows to accurately measure the LO phonon energy at 293.5 cm$^{-1}$, very close to the bulk value in GaAs.

**Figure 4**: Emission spectra of sample HM3872 for different values of the pump laser wavenumber. The scattered light strictly follows the injected one and it is always separated by one LO$_{ph}$. Note that the amount of scattered/emitted light is directly proportional to the absorption at the injection frequency.

**Figure 5:** Direct comparison between the emission spectra recorded under optical injection in two different semiconductor systems. The different value of the LO-phonon energy in GaAs and in InGaAs (lattice matched to InP) is clearly observable. To highlight the phonon separation in both system, the pump wavenumber was subtracted on the frequency axis.

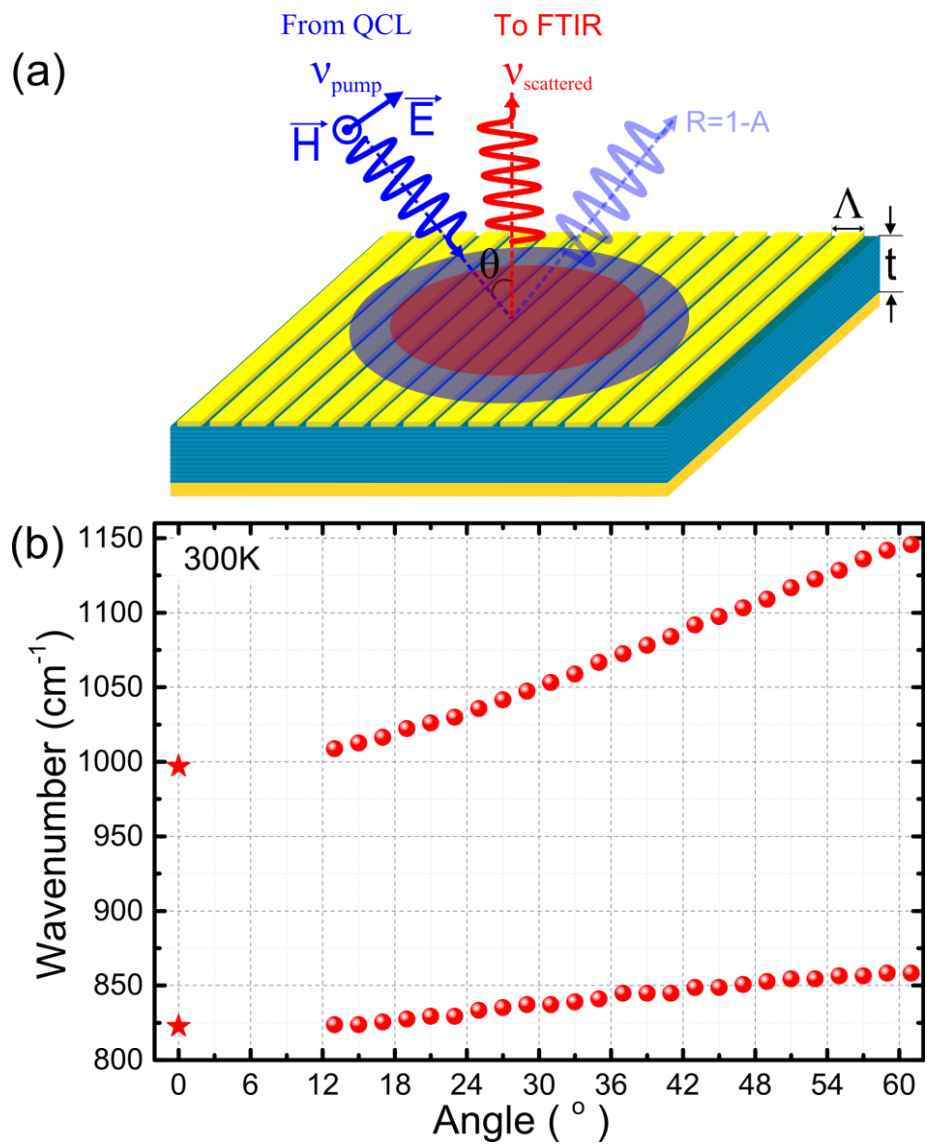

**Figure 1:** JM Manceau et al.

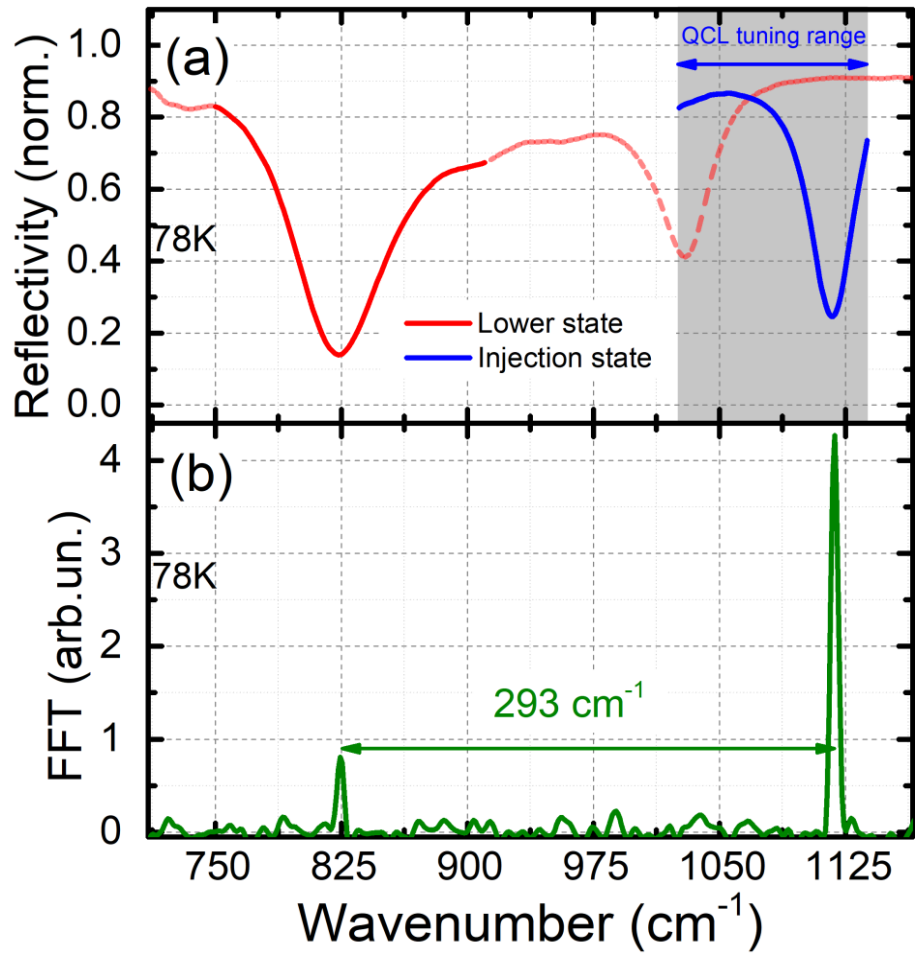

**Figure 2:** JM Manceau et al.

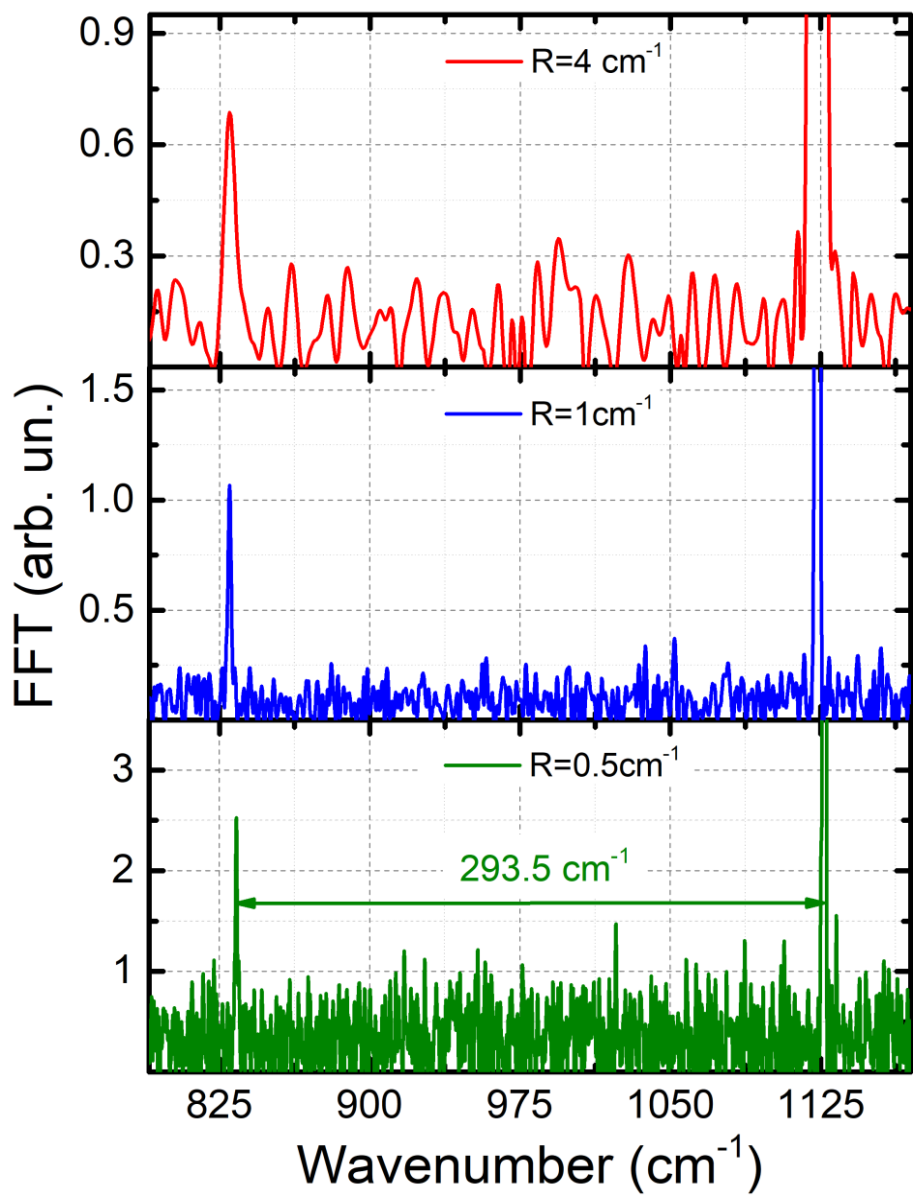

**Figure 3:** JM Manceau et al.

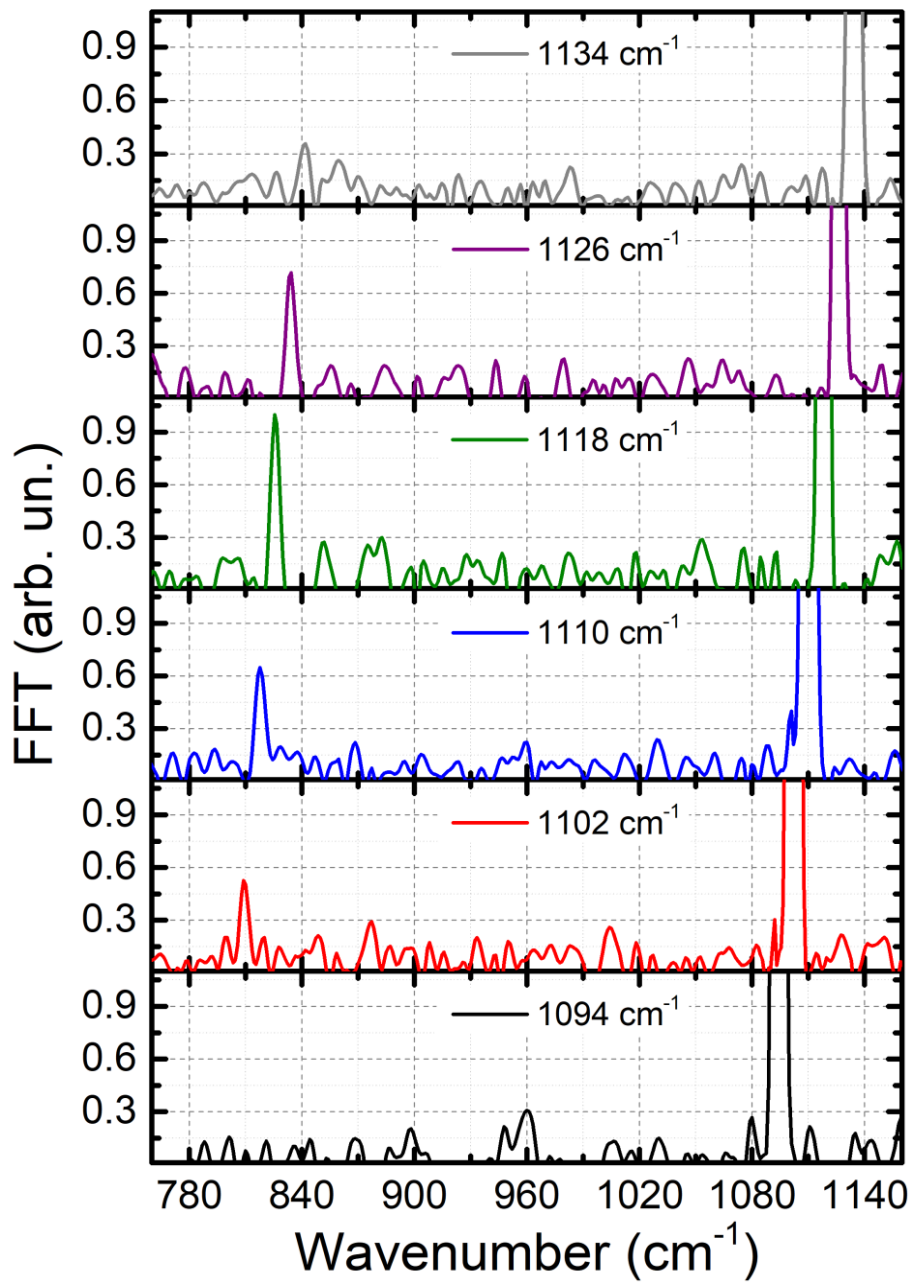

**Figure 4:** JM Manceau et al.

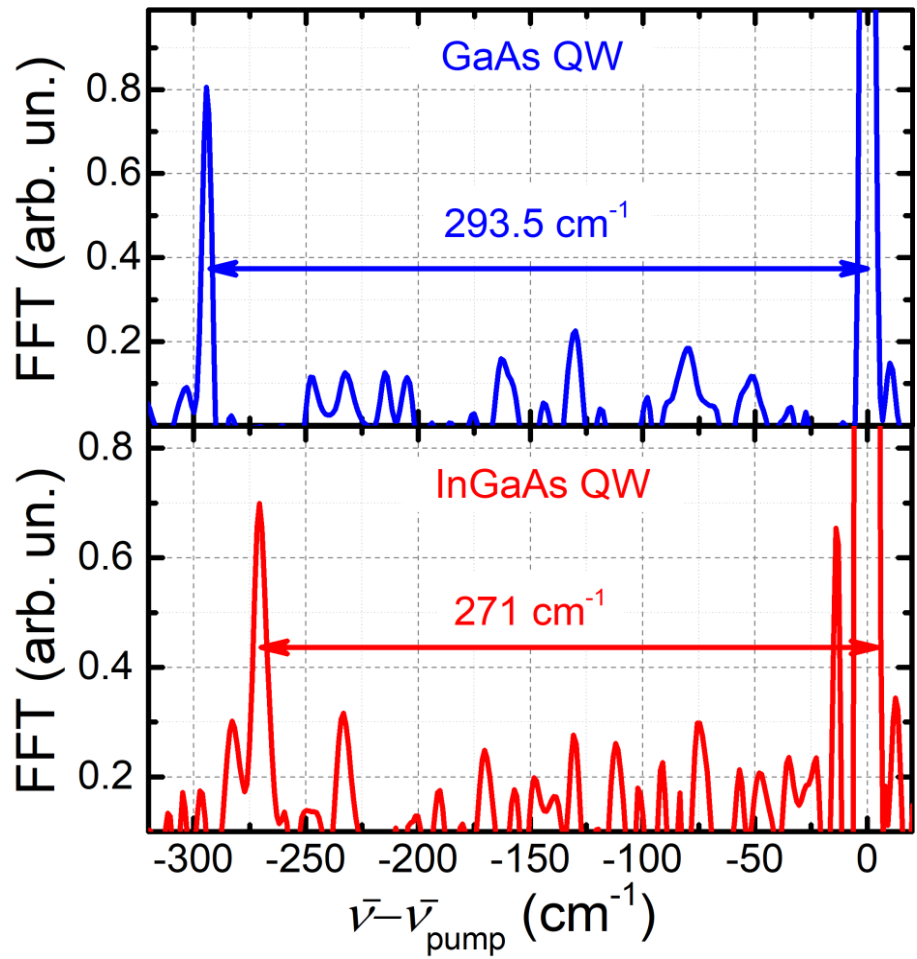

**Figure 5:** JM Manceau et al.